# Kepler-62f: *Kepler's* First Small Planet in the Habitable Zone, but Is It Real?

William Borucki[1], Susan E. Thompson[2], Eric Agol[3,2], and Christina Hedges[4]

[1]NASA Ames Research Center, Moffett Field, CA 94035, USA
[2]Space Telescope Science Institute, Baltimore, MD 21218, USA
[3]Department of Astronomy and Virtual Planetary Laboratory University of Washington, Seattle, WA 98195, USA
[4]Bay Area Environmental Research Institute, 625 2nd St Ste. 209, Petaluma, CA 94952

**Abstract**
Kepler-62f is the first exoplanet small enough to plausibly have a rocky composition orbiting within the habitable zone (HZ) discovered by the *Kepler Mission*. The planet is 1.4 times the size of the Earth and has an orbital period of 267 days. At the time of its discovery, it had the longest period of any small planet in the habitable zone of a multi-planet system. Because of its long period, only four transits were observed during *Kepler's* interval of observations. It was initially missed by the *Kepler* pipeline, but the first three transits were identified by an independent search by Eric Agol, and it was identified as a planet candidate in subsequent *Kepler* catalogs. However in the latest catalog of exoplanets (Thompson et al., 2018), it is labeled as a false positive. Recent exoplanet catalogues have evolved from subjective classification to automatic classifications of planet candidates by algorithms (such as 'Robovetter'). While exceptionally useful for producing a uniform catalogue, these algorithms sometimes misclassify planet candidates as a false positive, as is the case of Kepler-62f. In particularly valuable cases, i.e., when a small planet has been found orbiting in the habitable zone (HZ), it is important to conduct comprehensive analyses of the data and classification protocols to provide the best estimate of the true status of the detection. In this paper we conduct such analyses and show that Kepler-62f is a true planet and not a false positive. The table of stellar and planet properties has been updated based on GAIA results.

**Introduction**

The *Kepler Mission* was launched in 2009 with the goals of determining the occurrence rate of small (i.e., approximately Earth-size) planets in the habitable zones (HZ) of solar like stars and to determine the characteristics of exoplanets and the stars they orbit (Koch et al 2010). This mission was the culmination of five proposals to NASA, interspersed with years of tests to demonstrate the technology necessary for high-precision photometry (Borucki 2016; Borucki 2017). Quantitative discussions by Rosenblatt (1971), Borucki and Summers (1984) generated requirements for the technology needed for the photometric detection of Earth-size planets. Borucki and Koch (1994) compared various methods to observe thousands of stars simultaneously. Robinson et al. (1995) published the results of CCD tests that demonstrated CCD arrays had the 10 ppm relative precision needed when systematic errors were measured and corrected. The feasibility of doing automated photometry on thousands of stars simultaneously was shown by a wide-field-of-view photometric telescope that radioed its observations to Ames where the results were analyzed for transits by a computer program (Borucki et al. 2001). In the same year Koch et al (2001) published results from a laboratory facility that provided an end-to-end demonstration of prototype photometer that could reliably detect Earth-size transits in the presence of the noise expected on orbit. The fifth proposal to the NASA Discovery Program was approved for implementation in 2001 and the Mission launched in 2009. Dozens of candidate transiting planets were immediately obvious. The 2010/04/20 (#2) issue of Astrophysical Journal Letters was entirely devoted to the 24 papers describing the *Kepler* results. Caldwell et al. (2010) discuss *Kepler's* on orbit instrument performance. Jenkins et al, 2010 present an overview of the automated analysis of the *Kepler* data. Batalha et al. (2010) describe the process of choosing the best targets among the millions of stars in the field-of-view. Petigura et al. (2017) and Furlan et al. (2018) discuss the ground-based survey required to characterize the host stars. During its four-year life span, the Mission discovered more than 4400 planetary candidates with 2300 confirmed as planets, and of these, 13 were small planets in the HZ. *Kepler* revolutionized our understanding of exoplanet systems with short periods (Lissauer et al. 2014, Winn 2019), giving insight into their statistical properties, such as occurrence rates, size distributions, orbital periods, multiplicities, as well as turning up examples of new systems such as circumbinary transiting planets (Winn & Fabrycky 2015; Doyle 2019).

---

[1] Ames Associate
[2] Guggenheim Fellow

Determining the occurrence rate of planets requires a large, uniformly-processed sample whereas characterizing particular discoveries allows a more comprehensive and detailed examination of all the data available for each of the selected individuals.

The Kepler-62 system was one of the earliest confirmed exoplanet systems with planets in the HZ. The KOI that became 62e looked like our best small HZ planet candidate and was the initial focus of the effort but our detailed analysis revealed 62f was an even better candidate in both size and the amount of flux intercepted. Furthermore, Kepler-62f was a nearly Earth-size planet (~1.4 $R_\oplus$) and therefore more likely to have a rocky composition. At the time of its discovery, Kepler-62f was the smallest exoplanet in the HZ, making it a particularly important planet candidate. As such, a variety of different observation techniques were used to obtain the data necessary to establish its reality as a planet rather than a false-positive (Borucki et al. 2013). Infrared measurements were obtained from ground-based telescopes and the space-based Spitzer telescope. Telescopes using active optics, spectral masks, and speckle systems provided information on nearby stars that might cause confusion. Analysis of centroid motion of image centroids and a search for transit timing variations (TTVs) (Ragozzine & Holman 2019; Nesvory 2019) were conducted. High-resolution spectra from several ground-based telescopes were used to search for radial velocity (RV) variations, stellar characteristics, and variability. Upper limits on the masses were placed, based on the lack of TTV and RV variations, of <36 $M_\oplus$ for both 62e and 62f at 95% confidence (additional transits from later Quarters and from HST have not yet been included in this analysis). An extensive analysis of possible false-positive sources was conducted. Because of the resources required, this range of measurements and analyses are not conducted for most exoplanets. All these observations and analyses are consistent with Kepler-62f being a planet rather than a false-positive. However, in the recent exoplanet catalog (Thompson et al. 2018), Kepler-62f has been listed as a false-positive. We will show that this misclassification is due to particular circumstances associated with the transits of this planet and with the criteria used in the automatic vetting procedure of that catalog. We suggest methods to overcome such misclassifications.

The Kepler-62 system (KIC 9002278) was first designated as Kepler Objects of Interest (KOI) 701.01, 701.02, & 701.03 based on the first four months of data and all appear in the second catalog paper (Borucki et al. 2011 and online data). Further analysis showed the presence of two additional KOIs: 701.04 and 701.05. After further data and analyses led to their confirmation as planets, the designations were changed to Kepler-62 b, c, d, e, & f; consistent with their orbital periods.

To estimate the planetary parameters, a careful characterization of the stellar properties was necessary. Preliminary values given in Borucki et al. (2011) are shown in Table 1. The discovery of the two habitable-zone candidates, 62e and 62f, motivated further refinement of the stellar properties. The stellar parameters are critical for obtaining the planet radii, which are only measured as a ratio to the stellar radius from the transit depth. High SNR spectroscopic observations were used to derive an effective temperature, log(g), and metallicity of the star (Furlan et al. 2018). These values were then matched to stellar evolution models to estimate the stellar size, mass, luminosity and age. These values are given in columns 2 and 3 of Table 1. Analyses by Thompson et al., (2018), Berger et al. (2018), and Fulton and Petigura (2018) are provided for comparison in columns 4 through 6.

Based on the data from GAIA DR2 (Bailer-Jones et al. 2018; Luri et al. 2018; GAIA Collaboration 2018, we have also carried out an empirical measurement of the properties of the host star based upon the GAIA DR2 parallax, broad-band photometric measurements from SDSS, 2MASS and WISE, the density estimated from the period-duration relation of the five planets, and an empirically-calibrated mass-luminosity relation (Eker et al. 2015). The broadband fluxes along with the parallax and a small extinction correction give a precise luminosity of $0.2565\pm0.0045$ $L_\odot$. The mass-luminosity relation yields a mass estimate, while the density measurement from Borucki et al. (2013) then yields a radius estimate. Finally, the luminosity and radius together yield a temperature for the star. All of these parameters are discrepant with the measurements based upon spectroscopy and parallax from Berger et al. (2018) and Fulton & Petigura (2018) at the 2-sigma level. The origin of the discrepancy appears to lie in the density of the star which is inferred from the transit models; this should be reexamined in future photometric models of the light curve. These parameters are also listed in Table 1. The revised parameters for the star, anchored by the GAIA DR2 distance from Bailer-Jones et al. (2018), yield a slightly larger size for the star, and hence slightly larger size for both planets. The planet 62f remains below the 1.6-1.8 $R_\oplus$ cutoff for rocky planets (based upon measurements at shorter orbital period, Fulton et al. 2017), while 62e remains slightly above.

Table 1. Comparison of estimates of stellar and exoplanet properties.

| Publication | Borucki et al. (2011) | Borucki et al. (2013) | Thompson et al. (2018) | Berger et al. (2018) | Fulton & Petigura (2018) | This paper |
|---|---|---|---|---|---|---|
| $M_*(M_\odot)$ | 0.83 | 0.69±0.02 | | | 0.697-0.014+0.023 | 0.764±0.011 |
| $R_*(R_\odot)$ | 0.68 | 0.64±0.02 | 0.66±0.04 | 0.711±0.029 | 0.689±0.008 | 0.660±0.018 |
| $T_{eff}$ (K) | 4869 | 4925±70 | 4926±98 | 4859±97 | 4859±60 | 5062±71 |

| log(g [cm/s$^2$]) | 4.70 | 4.68±0.04 | | | | 4.683±0.023 |
|---|---|---|---|---|---|---|
| [Fe/H] | | -0.37±0.04 | | | -0.34±0.04 | |
| ρ [g/cc] | 3.73 | 3.8±0.3 | | | 2.88±0.23 | 3.8±0.3 |
| L∗ (L☉) | 0.23 | 0.21±0.02 | | 0.254±0.029 | 0.246±0.018 | 0.2565±0.0045 |
| Age [Gyr] | | 7±4 | | | 10$_{(10.01-0.25+0.10)}$ | |
| D [pc] | | 368 | | 300.87±1.2 | 300.87±1.2 | 300.87±1.2 |
| R$_{62e}$ [R$_⊕$] | | 1.61±0.05 | 1.72$_{-0.07}^{+0.10}$ | 1.84-0.08+0.10 | 1.827±0.050 | 1.670±0.051 |
| S$_{62e}$ [S$_⊕$] | | 1.2±0.2 | 1.24$_{-0.19}^{+0.27}$ | 1.352±0.028 | 1.3±0.1 | 1.321±0.098 |
| R$_{62f}$ [R$_⊕$] | | 1.41±0.07 | 1.43$_{-0.06}^{+0.08}$ | 1.531$_{-0.099}^{+0.084}$ | 1.501±0.066 | 1.461±0.070 |
| S$_{62f}$ [S$_⊕$] | | 0.41±0.05 | 0.44$_{-0.07}^{+0.09}$ | 0.477±0.010 | 0.5±0.04 | 0.466±0.034 |

(S$_⊕$ is the solar flux at the top of the Earth's atmosphere.)

To put Kepler-62 in context, we compare it with other small, temperate ("habitable-zone") planet candidates discovered with the *Kepler Mission*. Figure 1 shows the semi-major axes of these planets with radii less than 1.8 R$_⊕$ as a function of host-star temperature. The dimension of 1.6 R$_⊕$ is thought to be the approximate boundary between rocky and gas-rich planets at shorter orbital periods (Rogers 2015), while there appears to be a marked gap in planetary radius at short periods around 1.8 R$_⊕$ (Fulton et al. 2017).

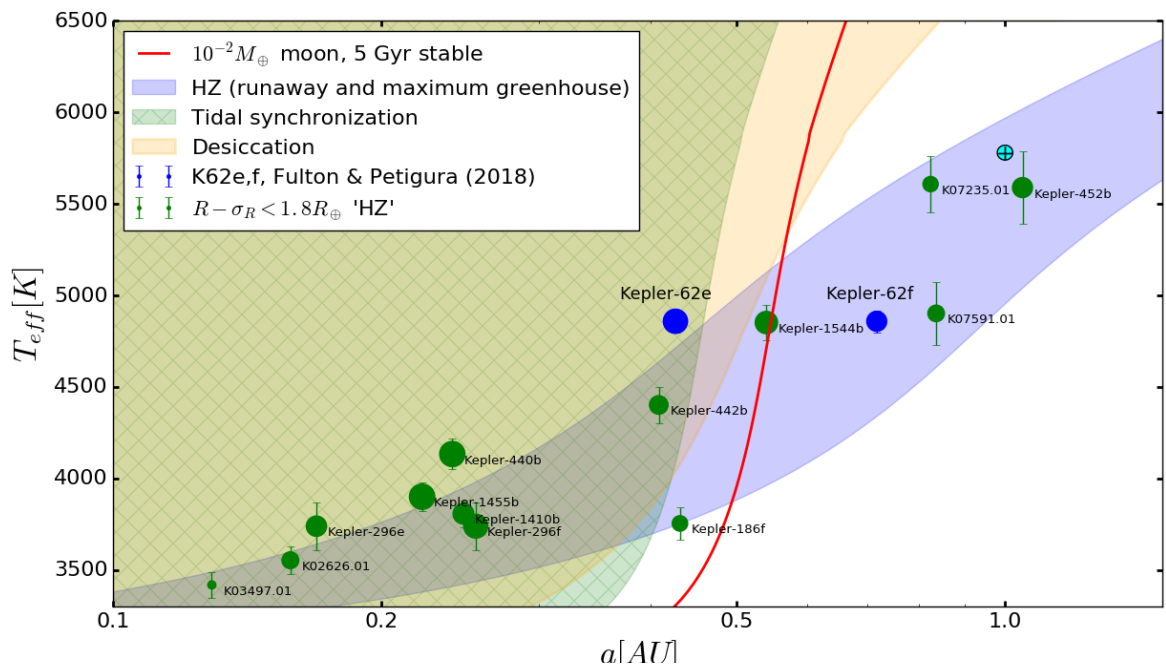

**Figure 1.** Stellar effective temperature versus semi-major axis for small planets (Rp < 1.8 R$_⊕$ to within 1-sigma) found with *Kepler* within and near the conservative "habitable-zone" (Kane et al. 2016), with modified parameters from Berger et al. (2018). The region shaded in blue/purple represents the "Conservative Habitable-zone"

boundaries as set by the runaway and maximum greenhouse limits (Kopparapu et al. 2014). Green represents the region where tidal synchronization (Kasting et al. 1993) is expected, while the orange region delineates the region of desiccation (Luger & Barnes 2015). The red line is the smallest value for the semi-major axis that would provide for dynamic stability for a 0.01 Earth-mass moon (Barnes & O'Brien 2002) for $Q_p=100$, $M_p=M_e$ and age 5 Gyr. Planets to the left and right of the blue HZ region are too hot and too cold, respectively, for liquid water to exist on their surface, given the atmospheric compositions assumed in the model. The portion of the purple region to the right of the colored regions is potentially the most favorable region for habitability; shaded regions are less favorable. The orbital distance and temperature of the host star are plotted in blue for Kepler-62e and 62f, based upon Fulton & Petigura (2018). The sizes of the planets in Figure 1 are based on the revised stellar radius utilizing the GAIA DR2 distance measurement based on the paper by Fulton & Petigura (2018). The sizes of the symbols for planets (relative to the Earth) are proportional to their actual size.

Overplotted are the runaway and maximum greenhouse limits corresponding to the inner and outer orbital distance at which a planet with an Earth-like atmosphere and carbon-silicate cycle might maintain surface temperatures supporting liquid water (Kopparapu et al. 2014). In addition to the incident stellar radiation, a star's gravity can also impact a planet's habitability (Barnes 2017). Planets which orbit at distances closer to the star will experience stronger tidal forcing. This can lead to tidal synchronization of the planet, slowing down its rotation, and the inability to host a giant moon. Such results negate two unique qualities of our Earth that have been argued to correlate with habitability (Balbus 2014), although these may not be required for habitability (Joshi 1997; Lissauer et al. 2012a).

The tidal synchronization timescale and moon instability timescale depend on a complex range of factors that are difficult to estimate for any particular system. Nevertheless, a rough estimate can be made for timescales of 5 Gyr for an Earth-like planet. The red curve in Figure1 shows the orbital distance at which an Earth-like moon would exit the Hill sphere and become unstable in 5 Gyr. The calculation assumes a tidal Q factor of 100 and Love number of 0.3 (Barnes & O'Brien 2002). The green-colored region in Figure 1 shows the range of distances for which a planet is expected to tidally synchronize its rotation in 5 Gyr based on a constant-time tidal lag model (Kasting et al. 1993). In order to be habitable in a way similar to that on Earth, the planet should orbit at a distance to the right of the green region. The true effect of tides on these planets, and the resulting impact on habitability, is likely to be much more complex, however, so these curves should only be viewed as a guide (Barnes 2017; Shields et al. 2016; Bolmont et al. 2015; Deitrick et al. 2018; Piro 2018; Sasaki & Barnes 2014).

Another limit to habitability is the ability of a planet to retain water during the period when the star is contracting to the main sequence. Late-type stars have extended pre-main sequence phases with elevated activity (Lissauer 2007). This activity is associated with high energy photons that irradiate the planet and can lead to water dissociation and escape of hydrogen, resulting in desiccation (Luger & Barnes 2015). In Figure 1 we plot the approximate boundary (i.e., to the right of the orange region) beyond which an Earth-like planet might avoid any significant water loss due to this process. Similarly, atmospheric or water loss may occur due to the higher orbital velocity within the habitable-zone for later type stars (Lissauer 2007; Zahnle & Catling 2017); this limit tends to affect only the latest-type stars, and so is not included in the plot. Of course, planets can migrate and water could be delivered or outgassed at later times, so this constraint has some latitude when arguing about potential habitability as defined by the presence of surface liquid water.

All of these constraints indicate that physical barriers to Earth-like habitability may occur for planets in the inner regions of "habitable-zones" of cooler, late-type stars. These stars have been the focus of transit searches because their planets have shorter orbital periods and a higher geometric probability of transit, and because these stars are especially common. All of these factors lead to an enhanced probability for transit detection, in addition to the smaller size of the star which leads to a larger depth of transit. In addition, the small stellar size improves the signal-to-noise ratio for the characterization of the planets' atmospheres via transit spectroscopy due to the larger depth of transit.

Kepler-62f is the first small transiting planet found by the *Kepler Mission* which has a sufficiently large orbital distance to experience weak tides and to avoid desiccation at its current location, making it a unique and particularly valuable planet at the time of its discovery. Since that time, *Kepler* added additional small planet candidates which also reside in this region (Kane et al. 2016). However, Kepler-62e and f are the only pair of temperate, small planets amongst these, which might enable the future characterization of these planets masses with transit-timing variations (Agol & Fabrycky 2017).

**A short history of the discovery of Kepler-62f**

After detection of the Kepler-62 system, additional ground-based and space-based observations and several analyses were used to confirm the Kepler-62 system as a real planetary system (Borucki et al., 2013). For example, Spitzer observations at 4.5 $\mu$m of Kepler 62-e by Désert et al. (2015) showed that the transit depth measured in the IR (i.e., 570±405 ppm) and visible (700±30 ppm) were similar as is expected if the transits are associated with the target star.

False positives in photometric searches for transiting planets are common, often caused by the presence of background eclipsing binaries. The fact that the Kepler-62 system shows transits from multiple objects greatly reduces the odds that it is a false-positive such as a background eclipsing binary (Lissauer et al. 2012b; Lissauer et al., 2014; Rowe et al, 2014). Mathematical analyses using the BLENDER algorithm (Torres et al. 2004; 2011; Fressin et al. 2012; Borucki et al. 2013; Torres, Kipping, Fressin, et al. 2015; Torres & Fressin 2019) estimate the odds that any of the five planets is a false positive rather than a planet is less than $1 \times 10^3$.

During the writing of the 2013 paper describing the discovery of the Kepler-62 system of three planets, the draft was sent to the *Kepler* team members for their review and additions. Concurrently, one of the team members (Eric Agol) had been working on an algorithm to detect exoplanets with significant transit timing variations. His approach used box-like transit models that were fit to every cadence in the dataset over a grid in transit duration and depth. Each result was compared to a fit with no transit, and the difference in the chi-square statistic ($\Delta\chi^2$) was computed between the fits with and without a transit. Results were fed into the QATS algorithm (Carter & Agol 2013, Agol & Carter 2019), that created a periodogram allowing for small fluctuations in the times of transit. This approach was developed in the search for circumbinary planets that were particularly problematic due to noisier light curves. The requirement of a consistent depth for all transits helped to suppress multiple noise fluctuations from masquerading as a series of transits to create a false transiting planet. With the standard QATS algorithm, a single strong noise feature can change the baseline level of noise in the periodogram, which makes the signal of real planets less significant. Requiring the same depth for all transits helped to clean the periodogram of this spurious noise, making the peaks in the periodogram of real planets more significant. This process was used to analyze the Kepler-62f light curve after the fitting and subtraction of the transit signatures from the three previously-known planetary candidates.

The results of the analysis showed three strong peaks in $\Delta\chi^2$ plotted versus time plot (Figure 2) with a period of 267.29 days. These peaks were so strong that the planet was readily recognized by eye when co-author Agol first viewed this plot; the QATS periodogram analysis was not needed to detect the planet. As shown in the Figure, the first transit of the planet is missing. The light curve was checked for an earlier transit and it was found to have occurred during a data gap when the spacecraft was downlinking data and not observing. Unfortunately the expected TTVs for Kepler-62f planets are of order 10s of minutes and therefore too small for detection when compared to the uncertainty of the center of the transit times as measured with *Kepler* (i.e., ~ 18 minutes).

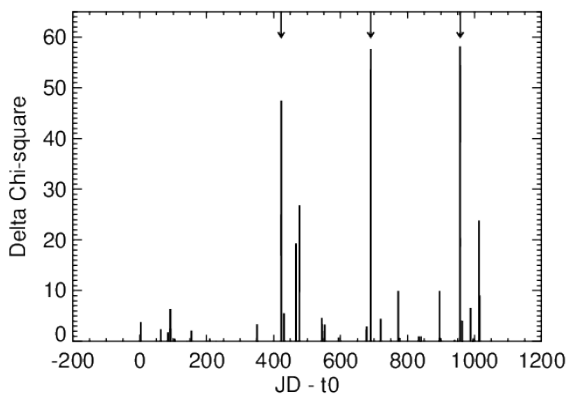

Figure 2. Three equally-spaced peaks in $\Delta\chi^2$ first demonstrated the presence of the Kepler-62f planet prior to the availability of the Q15 data. The lightcurve for Kepler-62 was preprocessed to remove the signal of the other 3 planets, leading to a strong, periodic signal from Kepler-62f. The vertical arrows show the locations of the three transits observed prior to the publication of the discovery paper (Borucki et al. 2013).

Shortly thereafter, another planet with a period of 12.4 days was found. The discovery of these additional planets caused the draft to be rewritten to present information on all five planets (Borucki et al., 2013). Later, when the Q15 data became available, a fourth transit of planet 62f was found at the predicted time, giving further confidence to the veracity of Kepler-62f.

The initial detection of Kepler-62f occurred at the end of August 2012 just after the Kepler data containing the third transit in Quarter 12 was released (Kepler Data Release 17). During the preparation of the paper, the data were made available on the PlanetHunters website[3]. After the paper was submitted (in January 2013), but before it was published (in April 2013), a citizen scientist, Mark Omohundro (@ajamyajax) identified the three transits by visual inspection, and, without knowledge of the *Kepler* paper under review, he reported his findings on the PlanetHunters website before the Kepler-62 paper was published.[4] Had Agol missed the detection of 62f, the credit for the first identification of the planet would have fallen to an amateur planet sleuth.

---

[3] https://www.planethunters.org
[4] https://oldtalk.planethunters.org/discussions/DPH101owp4

**Is Kepler-62f a false positive?**

Despite these observations, the last, uniformly-vetted *Kepler* exoplanet catalog (Data Release 25, Thompson et al. 2018) does not list Kepler-62f among planet candidates, although it does list Kepler-62b, c, d, & e. This is a surprise because the folded transit observations (Figure 3) clearly show a well-defined transit light curve. Its measured transit duration (7.46 ± 0.20 hours) is consistent with the measured orbital period and the estimated stellar diameter of 0.64 ± 0.02 $R_\circ$ for an orbital inclination of 89.90° ± 0.03°. An analysis of the likelihood of Kepler-62f being a false positive caused by an astrophysical event mimicking a planetary transit shows only a 1 in 5000 chance (Borucki et al. 2013, Supporting Online Material).

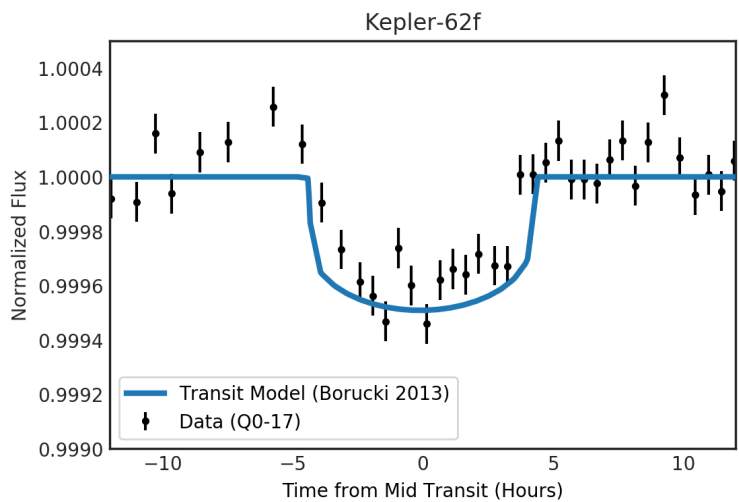

Figure 3. Phase-folded light curve for Kepler-62f for Quarters 0-17. Data from all 4 transits are averaged to produce the above light curve. One sigma errors are shown with vertical bars. The continuous curve shows the model fit from Borucki et al. 2013.

Catalogs prior to DR25 (Thompson et al. 2018) classified Kepler-62f as a KOI or planetary candidate. The first mention of what was eventually to become the Kepler-62f was based on the Q0 through Q2 data catalog (Borucki et al. 2011) where KOI 701.03 (later renamed as Kepler-62e) was listed (their Table 5) as a super-Earth planet in or near the HZ. KOI 701.01, -.02, and -.03 are listed in the Batalha et al. (2013) and Burke et al. (2014) catalogs. The Rowe et al. (2015) catalog based on the Q1-Q12 data was the first to list 701.04 (Kepler-62f) and 701.05(Kepler 62c). Kepler-62f is also listed in the Mullally et al. (2015) catalog. All five KOIs were detected and dispositioned as planet candidates in the DR24 catalog (Coughlin et al. 2016), the first uniform catalog with fully automatic vetting via the DR24 Robovetter. It is only in the most recent catalog (DR25) that Kepler-62f has been classified as a false positive.

The *Kepler* exoplanet catalogs evolved in order to accomplish *Kepler's* goal to determine the occurrence rate of small planets around other stars. Earlier catalogs (Borucki et al. 2011; Batalha et al. 2013; Burke et al. 2014; Rowe et al. 2015 and to some extent Mullally et al. 2015) manually vetted each system and only deemed signals "false positives" when there was strong evidence to that effect. The last *Kepler* catalog, Data Release 25 (DR25) (Thompson et al. 2018), used a robotic vetting algorithm called the Robovetter (Thompson et al. 2018). It also implemented new criteria (Thompson et al. 2018, Appendix A) that must be passed to be promoted to "planet candidate" in order to obtain a reliable catalog.

Automating the vetting had several advantages, including speed, flexibility and the ability to characterize which transits were incorrectly called "false positives". Simulated transits and false positives were processed using the Robovetter in order to find thresholds on each criterion. Each vetting parameter was tuned to maximize the recognition of simulated transits into "planet candidates" while minimizing the simulated false positives. Because all of the input signals to the Robovetter were treated in the same way, the detection system could be understood and its performance could be quantified for use in occurrence rate calculations. These measurements of the detection system are especially important near *Kepler's* detection threshold where the vetting algorithms are less reliable. This occurs at orbital periods greater than approximately 200 days and planet radii less than approximately 2 $R_\oplus$. Ultimately, the candidates near *Kepler's* detection threshold are the ones that will be used to estimate the occurrence rate of terrestrial planets in the HZ of GK dwarf stars (e.g. see Mulders et al. 2018).

This need for a consistent detection system for accurate occurrence rates also means that even in obvious cases of misclassification, manual overrides of its decisions were not allowed. While the Robovetter and the DR25 catalog are well suited for occurrence rates, it does not always provide the best knowledge of individual systems. Ultimately, failure to meet the Robovetter criteria for a "planet candidate" does not invalidate the existence of those planets confirmed by extensive observations and analysis. Indeed, measurements of the vetting system (Thompson et al. 2018) indicate that a significant fraction of real transit events were deemed "false positives" by the Robovetter in this part of the parameter space. The following discussion explains the situations that caused the Robovetter to disposition Kepler-62f as a false positive in the DR25 catalog.

Two Transit Events Rejected by Robovetter
One of the requirements of the DR25 Robovetter protocol is that the light curve show a minimum of three valid transits and that each transit pass several tests that check that it is consistent with that expected from a transiting planet. During *Kepler Mission* science operations, four transits of Kepler-62f were observed in Quarters 6, 9, 12, and 15 centered at times (BJD-2454833): 589.725, 857.006, 1124.287, and 1391.568. It is likely that another transit occurred at 322.434 during Quarter 3. However there are no observations at that time because the spacecraft was in a data-transmission mode rather than in a science acquisition mode. See Figure 4.

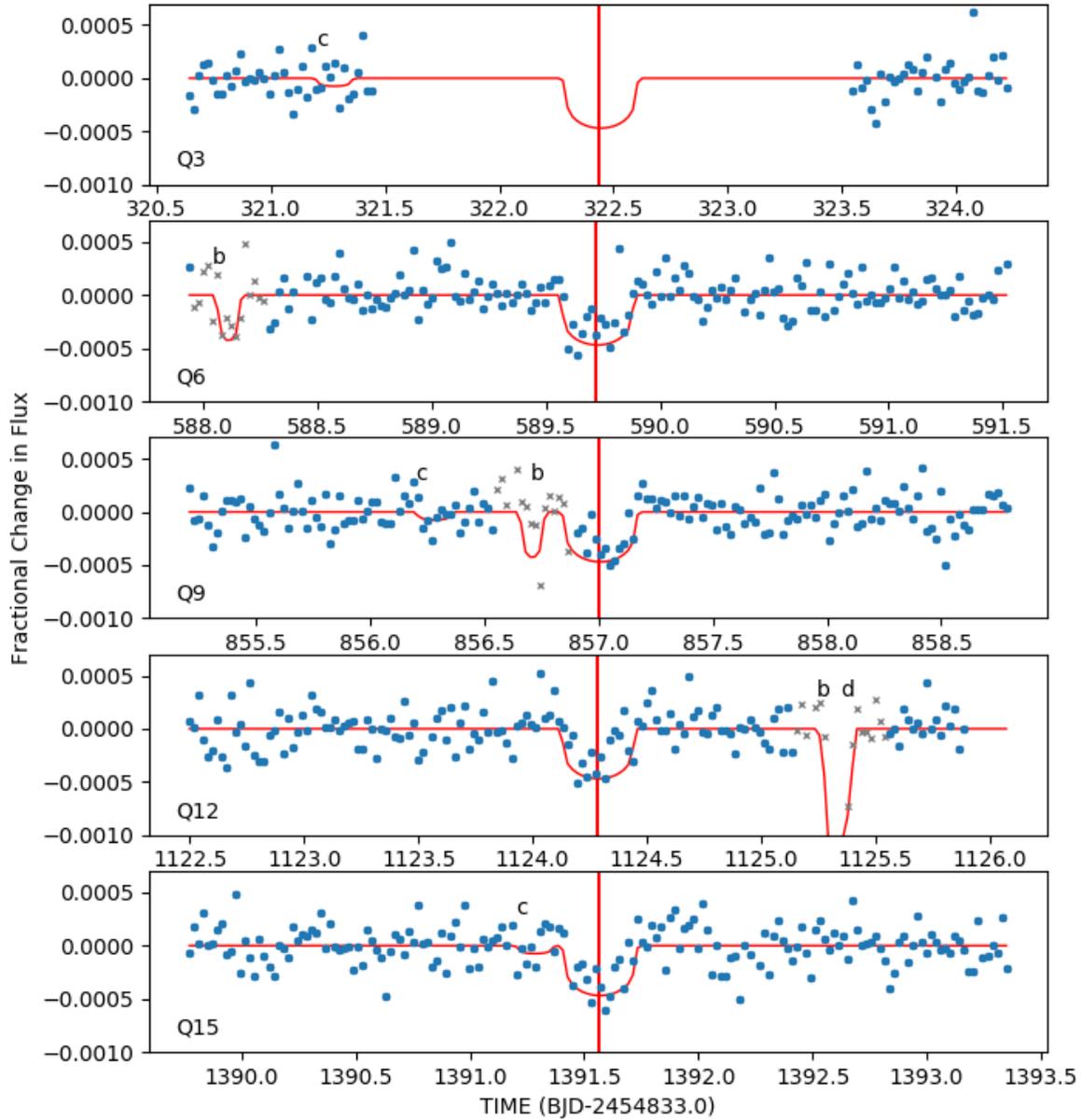

Figure 4. Observed fluxes (dots) of the five Kepler-62f transits that occurred during the *Kepler Mission*, centered on the mid-transit time for Kepler-62f and covering the two days before and after the mid-transit. Top to bottom; Quarters 3, 6, 9, 12, and 15. Note that the transits of Kepler-62f shown in panels 3 and 4 were not considered valid by the Robovetter. These data use the DR25 data processing and are median-detrended by the *Kepler* pipeline[5]. The

---

[5] The light curve shown corresponds to that in the MODEL_INT extensions of the Kepler Data Validation time series files available at the Mikulski Archive for Space Telescopes (doi:17909/T9CT1P).

red line shows the full transit model for Kepler-62b, c, d, e, & f as determined by the *Kepler* pipeline (Twicken, Jenkins, Seader, et al. 2016). Transits by other members of the Kepler-62 system are indicated by lower-case letters. Points shown as grey-colored "x" were removed after the detection of other planets in this systems and were not available to the detection pipeline nor the Robovetter when vetting Kepler-62f.

As shown in Figure 4, *Kepler* observed four transits of Kepler-62f. Panels 2 through 5 show clear evidence of transits with depths of approximately 460 ppm and durations of approximately 7.5 hours. However, two of these four transits failed to pass DR25 Robovetter tests ("Rubble" and "Skye"), namely the transits during Q9 and Q12. Because at least three transits are required for Robovetter to accept a candidate, Kepler-62f was labelled as a False Positive. Kepler-62b,c,d, & e continued to be recognized as candidates.

The test called "Rubble" checked that each reported transit contains at least ¾ of the expected number of data points given the duration of the transit and *Kepler's* regular cadence. For the transit in Quarter 9 (i.e., panel 3 in Figure 4), the *Kepler* pipeline removed the first few cadences (shown as grey "x"s) of the transit because those cadences overlapped with the transit of Kepler-62b. Since Kepler-62b was found prior to Kepler-62f and because the *Kepler* pipeline removes the cadences associated with transits before searching for another transit, those cadences were not available for the analysis of Kepler-62f. Thus, the Q9 transit was deemed to have insufficient data to be counted as a true transit by the Robovetter. However, a careful examination of the data prior to the removal of Kepler-62b shows sufficient data to validate the transit. Note, the shallow transits of planet c were found by the pipeline after planet f and this is why the proximity of planet c in Q15 did not impact Rubble's decision of this transit.

Another test, called "Skye", checked for cases where the transit times are coincident with other Threshold Crossing Events (i.e., potential transit signals found in the data) in the same "skygroup". (A skygroup includes all the targets observed on the same CCD detector during each Quarter.) When too many transits occur near the same time, it assumes that such signals are caused by instrumental systematics; specifically high-frequency oscillations often called "rolling band" artifacts (Caldwell et al. 2010) and illustrated in Figure 11 of Borucki (2016). However, real transits that occur at such times are likely to be inadvertently flagged as "bad". The Skye test failed the transit which occurred in Quarter 12; i.e., shown in panel 4 of Figure 4.

However, an analysis of the data taken at the time of the Q12 transit shows that the no "rolling band" artifact was present that could have mimicked a transit. The top panel of Figure 5 shows the flux time series from the 8 central illuminated pixels of the image of Kepler-62. The right panel shows the flux from the 22 pixels unilluminated (sky background) pixels that surrounded the 8 central pixels. Also shown is the expected flux reduction in the background pixels that would have occurred if a "rolling band" artifact had created the transit observed in the 8 illuminated pixels. Figure 5 shows that no dip in flux occurred in surrounding pixels. Therefore no "rolling band" artifact was present at the time of the Q12 transit. Thus the Q12 transit must be considered a legitimate transit.

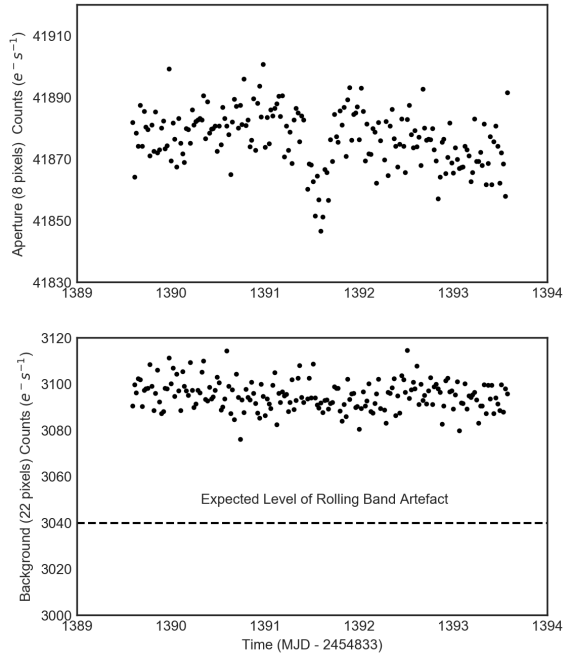

Figure 5. Light curve of Kepler-62 during and near the time of the Q12 transit. The upper panel displays the results for the sum of the count rate for the 8 pixels encompassing the star image and shows a prominent dip in flux at the expected time. The lower panel shows the sum for the surrounding, unilluminated 22 pixels at the same time. If a "rolling band" artifact had been present that caused the dip in the center pixels, the artifact would necessarily have reduced the flux in the 22 surrounding pixels to the level shown by the dotted line in the lower panel. The data show no such signal.

The uniform processing of the *Kepler* threshold crossing events by Robovetter produces a useful set of events for statistical analyses on the full *Kepler* exoplanet catalogue. However, the detailed analysis of Kepler-62f data shows that there are individual events that are misclassified. While both Rubble and Skye are useful tests to remove the majority of false positives, in the case of Kepler-62f they have removed a valuable small exoplanet in the HZ from the *Kepler* catalogue.

These results suggest that the detection of long-period planets would be enhanced if their search were conducted; 1) with a method that did not disqualify their transits when they occurred near the time of the transits of the short-period planets, and 2) by comparing the central image pixels with the surrounding pixels when the transit occurs during periods of high-level noise.

**Possibility that Kepler-62f is a false alarm due to random noise**
The reliability of the DR25 candidates (against instrument noise sources) has been measured (Thompson et al., 2018). The results have implications for validated planets like Kepler-62f. In particular, the validation of these planets is contingent on the observed change in the flux being caused by one of the astrophysical signals considered in the validation (like a background eclipsing binary). However, if there is significant probability that the observed transit signal is caused by random fluctuations in the noise level rather than by particular types of instrument noise, then the statistical estimate is not valid because this scenario is not included in the likelihood function (Mullally et al. 2018). Although Kepler-62f presents four transits and has a high (~12) SNR signal, there is always some possibility that Kepler-62f is a false-alarm. Only with additional observational evidence of the planet's existence can this possibility be further reduced. To this end, Kepler-62f was observed with the Hubble Space Telescope on 2017/12/7-8 (Program 15129, Chris Burke PI) with the express purpose of validating the planet transits. Preliminary results indicate a transit was observed at the expected time.

**Summary**

This short history of the discovery of Kepler-62f points out the usefulness of detailed examinations of exoplanet false positives by a variety of approaches. This is particularly important for the validation of small, long-period planets when few transits are available for analysis. Automated classification systems (such as Robovetter) provide uniformly vetted catalogs of exoplanet systems which is crucially needed for robust statistical analysis. However such systems can miss events when the data do not conform to simplified criteria. In particular, the classification system (Robovetter) used to produce the DR25 catalog ignored two of the four transits of Kepler-62f and subsequently classified it as a false-positive because it required a minimum of three acceptable transits. This paper presented a comprehensive examination of the time-series data for two transits that were not recognized by Robovetter.

The transit occurring during the Quarter 9 observations took place near the transit time of another planet in the system. That propinquity caused some of its data points to be ignored. The transit occurring during the Q12 period was ignored by Robovetter because of the possibility that a "rolling band" artifact might be responsible for the decreased signal. Comparing the transit depth of the central pixels used to measure the star brightness with those of the surrounding pixels demonstrated that the artifact could not have been the source of the transit. Therefore the four expected transits of Kepler-62f were present and valid.

This is presented as an example of how uniform processing using data only from *Kepler*, while necessary for statistical studies, is not well suited for making determinations of individual objects. It is a mistake to consider the DR25 catalog disposition reason to cast doubt on the confirmation of Kepler-62f, which used data beyond that directly available from *Kepler* system. The analysis presented here are consistent with the conclusions in Borucki et al., 2013; i.e., that Kepler-62f should be considered a confirmed planet.

Given the likely veracity of this planet, it remains one of the most interesting habitable-zone planets found to date. The longer period of this planet may make it possible for it to avoid desiccation, tidal synchronization and moon loss, all characteristics which may relate to habitability. Several bars of $CO_2$ should be sufficient to allow for surface liquid water, if the planet hosts an Earth-like atmosphere (Shields et al. 2016). Unfortunately, due to the faintness of the host star, atmospheric transmission spectroscopy will not be feasible with JWST. However, given these intriguing prospects for habitability, Kepler-62f and its companion 62e will continue to be interesting targets for characterization, perhaps with transit-timing variations.

A table of updated stellar and exoplanet properties that includes corrections based on recent GAIA results is also included.

**Acknowledgements**

Funding for the *Kepler Mission* was provided by NASA's Science Mission Directorate. Eric Agol acknowledges NSF grants AST-0645416 and AST-1615315, and NASA grant NNX13AF62G. Support for this work was provided by NASA through an award issued by JPL/Caltech. The authors would like to acknowledge the helpful suggestions and valuable discussions from Jeffrey L. Coughlin and Jack Lissauer and the anonymous reviewer

**References**

Agol, E. & Fabrycky, D.C. 2017. Transit-Timing and Duration Variations for the Discovery and Characterization of Exoplanets. Handbook of Exoplanets, Edited by Hans J. Deeg and Juan Antonio Belmonte. Springer Living Reference Work, ISBN: 978-3-319-30648-3, 2017, id.7

Agol, E. & Carter, J. A., 2019. Discovery and Characterization of Kepler-36b, *New Astronomy Reviews*; Special Issue on Kepler Exoplanet Firsts (in this issue).

Bailer-Jones, C.A.L., Rybizki, J., Fouesneau, M., Mantelet, G., Andrae, R. 2018. Estimating Distance from Parallaxes. IV. Distances to 1.33 Billion Stars in Gaia Data Release 2. *AJ* **156**: 58.

Balbus SA. 2014, Dynamical, biological and anthropic consequences of equal lunar and solar angular radii. *Proc. R. Soc. A* **470**: 20140263.

Barnes, J.W. & O'Brien, D.P. 2002. Stability of Satellites around Close-in Extrasolar Giant Planets. *ApJ* **575**:1087-1093.

Barnes, R. 2017. Tidal locking of habitable exoplanets. *Celestial Mechanics and Dynamical Astronomy* **129**:509-536.

Batalha, N.M., Borucki, W.J., Koch, D.G., et al., 2010, Selection, prioritization, and characteristics of *Kepler* target stars. *ApJL* **713**:L109-114.

Batalha, N. M., Rowe, J. F., Bryson, S. T., et al. 2013, Planetary candidates observed by *Kepler*. III. Analysis of the first 16 months of data. *ApJS* **204**:24.

Berger, T.A., Huber, D., Gaidos, E., & van Saders, J.L. 2018, Revised Radii of *Kepler* Stars and Planets using Gaia Data Release 2, arXiv:1805.00231

Bolmont, E., Raymond, S. N., Leconte, J., Hersant, F., & Correia, A.C.M. 2015. Mercury-T: A new code to study tidally evolving multi-planet systems. Applications to Kepler-62. *A&A* **583**:116.

Borucki, W.J., and Summers, A.L., 1984, The photometric method of detecting other planetary systems, *Icarus* **58**: 121.

Borucki, W., Koch, D G., Basri, G., et al. 2011, Characteristics of planetary candidates observed by *Kepler*. II. Analysis of the first four months of data. *ApJ* **736**:19.

Borucki, W., Agol, E., Fressin, F., et al. 2013, Kepler-62: A five-planet system with planets of 1.4 and 1.6 Earth radii in the habitable zone. *Science* **340**, 587-590.

Borucki, W. J., 2016, *Kepler Mission*: Development and Overview. *Reports on Progress in Physics* **79**, 036901.

Borucki, W.J., 2017, *Kepler*: A Brief Discussion of the Mission and Exoplanet Results, *Proceedings of the American Philosophical Society*, **161**:38

Burke, C. J. Bryson S. T., Mullally, F., et al. 2014, Planetary Candidates observed by *Kepler* IV: Planet sample from Q1-Q8 (22 months). *ApJS* **210**:19.

Caldwell, D. A., Kolodziejczak, J. J., Van Cleve, J. E., et al., 2010, Instrument performance in *Kepler's* first months. *ApJL* **713**:L92-96.

Carter, J. A. & Agol, E. 2013, The Quasiperiodic Automated Transit Search algorithm. *ApJ* **765**:1322.
Coughlin, J. L., Mullally, F., Thompson, S. E., et al. 2016, Planetary candidates observed by *Kepler*. VII. The first fully uniform catalog based on the entire 48-month data set (Q1-Q17 DR24). *ApJS* **224**, 12.
Deitrick, R., Barnes, R., Quinn, T.R., Armstrong, J., Charnay, B., and Wilhelm, C. 2018. Exo-Milankovitch Cycles. I. Orbits and Rotation States. *AJ* **155**:60.
Désert, J. M., Charbonneau, C., Torres, G., et al. 2015, Low false positive rate of *Kepler* candidates estimated from a combination of Spitzer and follow-up observations. *ApJ* **804**:59.
Doyle, L., 2019, The Discovery of "Tatooine": Kepler-16b *New Astronomy Reviews*; Special Issue on Kepler Exoplanet Firsts (in this issue).
Fressin, F., Torres, G., Rowe, J. F., et al. 2012, Two Earth-sized planets orbiting Kepelr-20. *Nature* **482**, 195.
Fulton, B.J., Petigura, E.A., Howard, A.W., Isaacson, H., Marcy, G.W., Cargile, P.A., Hebb, L., Weiss, L.M., Johnson, J.A., Morton, T.D., Sinukoff, E., Crossfield, I.J.M. and Hirsch, L.A., 2017 *AJ* **154** 109.
Fulton, B.J., & Petigura, E.A. 2018, The California Kepler Survey VII. Precise Planet Radii Leveraging Gaia DR2 Reveal the Stellar Mass Dependence of the Planet Radius Gap. arXiv:1805.01453.
Furlan, E., Ciardi, D.R., Cochran, W.D., et al., 2018. The *Kepler* Follow-up Observation Program; II. Stellar parameters from medium- and high-resolution spectroscopy. ApJ 861:149.
Gaia Collaboration, Brown, A.G.A., Vallenari, A., Prusti, T., de Bruijne, J.H.J., Babusiaux, C., Bailer-Jones, C.A.L. 2018. Gaia Data Release 2. Summary of the contents and survey properties. arXiv:1804.09365.
Jenkins, J. M., Caldwell, D. A., Chandrasekaran, H., et al., 2010, Overview of the Kepler Science Processing Pipeline, *ApJL* **713**:L87-91.
Joshi, M.M., Haberle, R.M., & Reynolds, R.T. 1997, Simulations of the Atmospheres of Synchronously Rotating Terrestrial Planets Orbiting M Dwarfs: Conditions for Atmospheric Collapse and the Implications for Habitability. *Icarus*, **129**: 450
Kane, S.R., et al. 2016. A Catalog of *Kepler* Habitable Zone Exoplanet Candidates. *ApJ* **830**:1
Kasting, J.F., Whitmire, D.P., and Reynolds, R.T. 1993. Habitable Zones around Main Sequence Stars. *Icarus* **101**:108-128.
Koch, D.G., et al., CCD photometry tests for a mission to detect Earth-sized planets in the extended solar neighborhood, 2000. *SPIE* **4013**:508K.
Koch, D. G., Borucki, W. J., Basri, G., Batalha, N. M., Brown, T. M. et al., 2010, *Kepler Mission* design, realized photometric performance, and early science., *ApJL* **713**:L79-L86.
Kopparapu, R.K., Ramirez, R.-M., Schottel-Kotte, J., Kasting, J.F., Domagal-Goldman, S., & Eymet, V. 2014. Habitable Zones around Main-sequence Stars: Dependence on Planetary Mass. *ApJL* **787**:L29.
Lissauer, J. J., 2007. Planets Formed in Habitable Zones of M Dwarf Stars Probably Are Deficient in Volatiles. *ApJ* **660**: L149.
Lissauer, J. J., Barnes, J. W., & Chambers, J.E., 2012a, Obliquity variations of a moonless Earth. *Icarus*, **217**: 77
Lissauer, J.J., and 23 co-authors, 2012b. Almost All of *Kepler's* Candidate Multiple Planet Candidates are Planets. Astrophys. J. 750, 112 (15pp).]

Lissauer, J.~J., Dawson, R.~I., \& Tremaine, S.\ 2014, Advances in exoplanet science from *Kepler*, *Nature*, **513**:336

Lissauer, J. J., Marcy, G.W., Bryson, S.T., et al. 2014, Validation of *Kepler's* Multiple Planet Candidates. II. Refined Statistical Framework and Descriptions of Systems of Special Interest. *ApJ* **784**:44.

Luger, R. & Barnes, R. 2015. Extreme Water Loss and Abiotic Oxygen Buildup on Planets Throughout the Habitable Zones of M Dwarfs. *Astrobiology* **15**:119-143

Luri, X., Brown, A.G.A., Sarro, L.M., Arenou, F., Bailer-Jones, C.A.L., Castro-Ginard, A., de Bruijne, J., Prusti, T., Babusiaux, C., Delgado, H.E. 2018. Gaia Data Release 2: using Gaia parallaxes. arXiv:1804.09376. Mullally, F, Coughlin, J. L., Thompson, S. E., et al. 2015, Planetary candidates observed by *Kepler*. VI. Planet sample from Q1—Q16 (47 months). *ApJS* **217**:31.

Mullally, Fergal; Thompson, Susan E.; Coughlin, Jeffery L.; et al. 2018, *Kepler's* Earth-like planets should not be confirmed without independent detection: The case of Kepler-452b, arXiv180311307M.

Nesvorny, D. 2019, How to Find a Planet from Transit Variations, *New Astronomy Reviews*; Special Issue on Kepler Exoplanet Firsts (in this issue).

Petigura, E.A., Howard, A.W., Marcy, G.W., et al., 2017, The California-Kepler Survey. I. High-resolution spectroscopy of 1305 stars hosting *Kepler* transiting planets. *AJ* **154**:107.

Piro, A.L. 2018. Exoplanets Torqued by the Combined Tides of a Moon and Parent Star. arXiv:1803.01971.

Ragozzine,D., & Holman, M. 2019, Kepler-9: the First Multi-Transiting System and the First Transit Timing Variations, *New Astronomy Reviews*; Special Issue on Kepler Exoplanet Firsts (in this issue).

Robinson, L. B., Wei., M. Z., Borucki, W. J., et al. 1995, Test of CCD precision limits for differential photometry, PASP 107:1094.

Rogers, L.A. 2015. Most 1.6 Earth-radius Planets are Not Rocky. *ApJ* **801**:41.

Rosenblatt, F., 1971, A two-color photometric method for detection of extra solar planetary systems, *Icarus* **14** 71R.

Rowe, J. F., Bryson, S. T., Marcy, G. W., et al. 2014, Validation of *Kepler's* multiple planet candidates. III. Light curve analysis and announcement of hundreds of new multi-planet systems. *ApJ* **784**:45.

Rowe, J F., Coughlin J. L., Antoci, V., et al. 2015, Planetary candidates observed by *Kepler*. V. Planet sample from Q1-Q12 (36 months), *ApJS* **217**:16.

Sasaki, T. & Barnes, J.W. 2014. Longevity of moons around habitable planets. *International Journal of Astrobiology* **13**:324-336.

Shields, A.L., Barnes, R., Agol, E., Charnay, B., Bitz, C., & Meadows, V.S. 2016. The Effect of Orbital Configuration on the Possible Climates and Habitability of Kepler-62f. *Astrobiology* **16**:443-464.

Thompson, S. E., Coughlin, J. L., Hoffman, K., et al. 2018, Planetary candidates observed by *Kepler*. VIII. A fully automated catalog with measured completeness and reliability based on Data Release 25. ApJS 235:38.

Torres, G., Konacki, M., Sasselov, D. D., & Jha, S., 2004, New data and improved parameters for the extrasolar transiting planet OGLE-TR-56b *ApJ* **614**:979.

Torres, G., Fressin, F., Batalha, N. M., et al. 2011, Modeling *Kepler* transit light curves as false positives: Rejection of blend scenarios for Kepler-9, and validation of Kepler-9d, a super-Earth-size planet in a multiple system. *ApJ* **727**, 24.

Torres, G., Kipping, D. M., Fressin, F., et al., 2015, Validation of 12 small *Kepler* transiting planets in the habitable zone. *ApJ* **800:**99

Torres, G., & Fressin, F., 2019, Discovery of the first Earth-sized planets in the Kepler-20 system, *New Astronomy Reviews*; Special Issue on Kepler Exoplanet Firsts (in this issue).

Twicken, J. D., Jenkins, J. M., Seader, S. E., et al., 2016, Detection of potential transit signals in 17 Quarters of *Kepler* data: Result of the final Kepler Mission Transiting Planet Search (DR25). *AJ* **152**: 158.

Winn, J.N., & Fabrycky, D.C., 2015, The Occurrence and Architecture of Exoplanetary Systems, *Annual Reviews of Astronomy and Astrophysics*, **53**:409

Winn, J. N., Kepler-78 and the Ultra-Short-Period Planets, *New Astronomy Reviews*; Special Issue on Kepler Exoplanet Firsts (in this issue).

Zahnle, K.J., Catling, D.C. 2017. The Cosmic Shoreline: The Evidence that Escape Determines which Planets Have Atmospheres, and what this May Mean for Proxima Centauri B. The *AJ* **843**: 122.